\documentclass[aps,pra,onecolumn,superscriptaddress]{revtex4-2}
\pdfoutput=1 

\usepackage[percent]{overpic}
\usepackage{amsmath}
\usepackage{siunitx}
\usepackage{graphicx}
\usepackage{subcaption}
\usepackage{makecell}
\usepackage{booktabs}
\usepackage{textcomp}
\usepackage{tablefootnote}

\begin{document}

\title{A Fibre-Integrated SPDC Heralded Single-Photon Source Using Zn-Indiffused PPLN Ridge Waveguides at Telecom Wavelengths}

\author{Peter Iveson}
\email{pwi1a19@soton.ac.uk}
\affiliation{Optoelectronics Research Centre, University of Southampton, University Road, Southampton SO17 1BJ, UK}

\author{Paolo L.~Mennea}
\affiliation{Optoelectronics Research Centre, University of Southampton, University Road, Southampton SO17 1BJ, UK}

\author{Goronwy Tawy}
\affiliation{Optoelectronics Research Centre, University of Southampton, University Road, Southampton SO17 1BJ, UK}

\author{Rex H.~S.~Bannerman}
\affiliation{Optoelectronics Research Centre, University of Southampton, University Road, Southampton SO17 1BJ, UK}

\author{Noe Palomar-Davidson}
\affiliation{Optoelectronics Research Centre, University of Southampton, University Road, Southampton SO17 1BJ, UK}

\author{Lewis D.~Wright}
\affiliation{Covesion Ltd., Unit F3, Adanac Park, Nursling, Southampton SO16 0BT, UK}

\author{Patrick M. Ledingham}
\affiliation{Optoelectronics Research Centre, University of Southampton, University Road, Southampton SO17 1BJ, UK}

\author{Peter G.~R.~Smith}
\affiliation{Optoelectronics Research Centre, University of Southampton, University Road, Southampton SO17 1BJ, UK}

\author{James C.~Gates}
\affiliation{Optoelectronics Research Centre, University of Southampton, University Road, Southampton SO17 1BJ, UK}

\author{Corin B.~E.~Gawith}
\affiliation{Optoelectronics Research Centre, University of Southampton, University Road, Southampton SO17 1BJ, UK}
\affiliation{Covesion Ltd., Unit F3, Adanac Park, Nursling, Southampton SO16 0BT, UK}

\begin{abstract}
We report a fibre-integrated heralded single-photon source based on a commercial Zn-indiffused MgO:PPLN ridge waveguide operated in Type-0 degenerate spontaneous parametric down-conversion (SPDC). The device emits a broadband spectrum spanning approximately 70~nm across the telecom C-and L-bands, with an absolute brightness of $9.1\times 10^{9}~\mathrm{pairs\,s^{-1}\,mW^{-1}}$. After accounting for optical losses, we determine an internal heralding efficiency of \SI{58(5)}{\percent}. The source maintains high purity, exhibiting a coincidences-to-accidentals ratio exceeding $9\times 10^{4}$ and a heralded second-order correlation of $g_{\mathrm{h}}^{(2)}(0) = (5.53 \pm 0.46)\times 10^{-4}$. Low multi-photon noise and broadband emission demonstrate the viability of such platforms for wavelength-multiplexed quantum networking.
\end{abstract}

\maketitle

\section{Introduction}

Heralded single-photon sources based on the second-order nonlinear process of spontaneous parametric down-conversion (SPDC) underpin a range of quantum technologies including quantum communications\cite{RNTanzilliQC}, quantum key distribution\cite{RNQKD}, and quantum metrology\cite{RNAlibart2016}. Lithium niobate (LiNbO$_3$) remains a leading platform due to its wide transparency window\cite{Berry2019}, large nonlinear coefficient\cite{RNWeisGaylord1985}, and compatibility with quasi-phase matching through periodic poling. Waveguide implementations enhance nonlinear interaction strength, enable stable spatial mode confinement, and facilitate efficient fibre coupling\cite{RNRidge}. Zn-indiffused ridge waveguides in particular offer low propagation loss, robust fabrication, and reduced photorefractive sensitivity\cite{RNYoungPRD1991}.

Recent demonstrations based on Zn‑indiffused ridge waveguides have shown these platforms to be efficient SPDC sources~\cite{Yadav2022,Bock2016}. However, such studies have typically focused either on correlated photon-pair performance metrics in fibre-integrated systems~\cite{Yadav2022}, or on heralded single-photon characteristics in free-space implementations~\cite{Bock2016}. To the best of our knowledge, a comprehensive quantum characterisation of fibre-integrated Zn‑indiffused ridge waveguides --- encompassing brightness, heralding efficiency, multi-photon suppression, and broadband spectral behaviour --- together with quantitative comparison to waveguide-based theoretical models, has not yet been fully reported.

\begin{figure}[t]
  \centering
  \begin{subfigure}[t]{0.48\linewidth}
    \centering
    \caption*{(a)}
    \includegraphics[width=\linewidth]{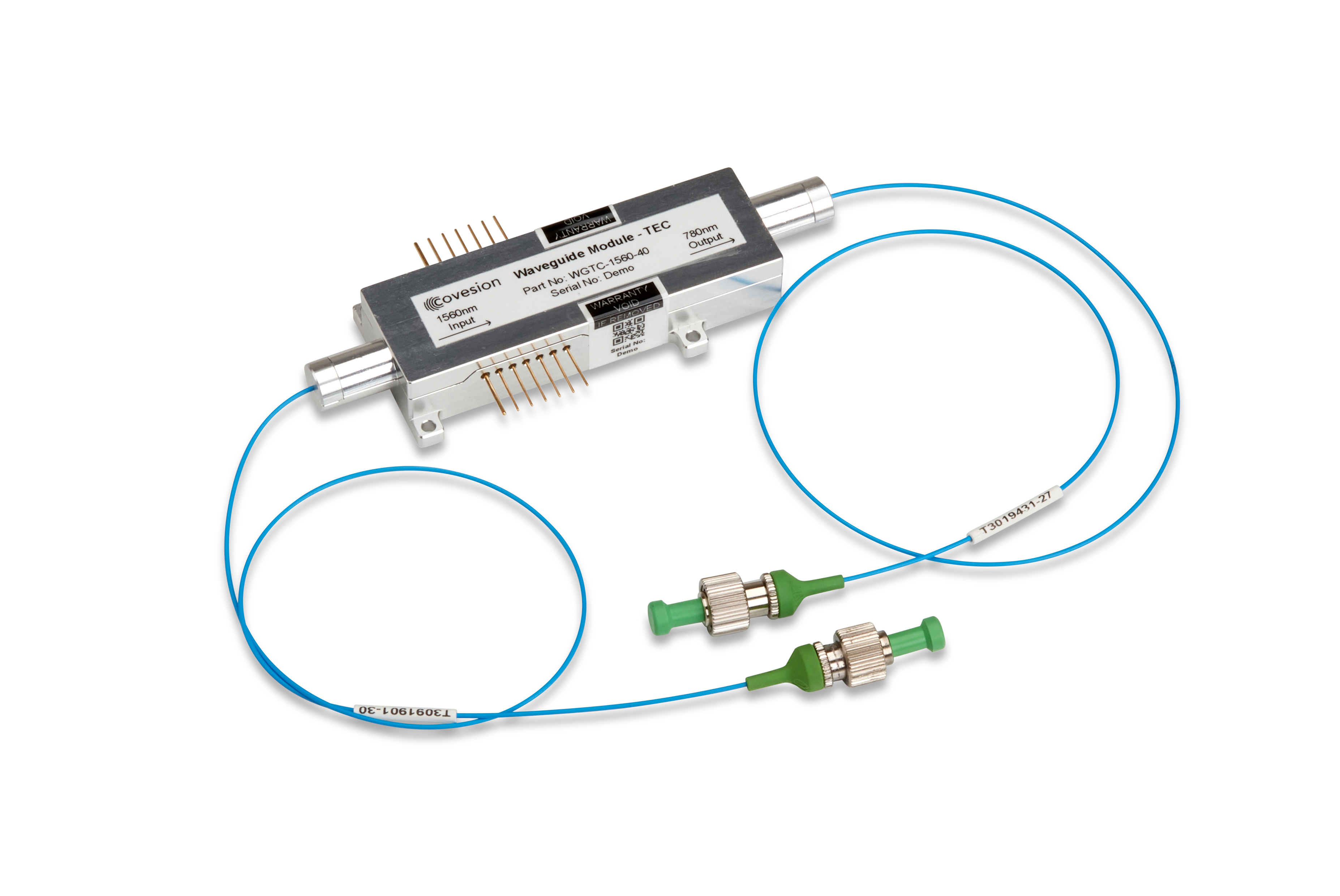}
    \label{fig:waveguidemodule:a}
  \end{subfigure}\hfill
  \begin{subfigure}[t]{0.48\linewidth}
    \centering
    \caption*{(b)}
    \includegraphics[width=\linewidth]{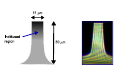}
    \label{fig:waveguidemodule:b}
  \end{subfigure}
  \caption{(a) Covesion Ltd. fibre-pigtailed waveguide module (Figure courtesy of Covesion Ltd.). (b) Left: Schematic of a Zn-indiffused ridge waveguide. Right: Micrograph of representative diced ridge facet}
   \label{fig:waveguidemodule}
\end{figure}

In this work, we characterise a fibre-pigtailed MgO:PPLN ridge waveguide (Fig.~\ref{fig:waveguidemodule}) operated in the Type-0 ($e \rightarrow e + e$) SPDC regime at $780~\mathrm{nm} \rightarrow 1560~\mathrm{nm}$. We combine experimental measurements with calibrated numerical modelling based on prism-coupled effective index data, enabling accurate simulation of modal dispersion, nonlinear overlap, and photon-pair generation rates. Using superconducting nanowire single-photon detectors (SNSPDs), we observe broadband emission spanning approximately 70~nm, together with high internal heralding efficiency and very low multi-photon noise. These results provide a detailed experimental and theoretical benchmark for a commercially available fibre-integrated SPDC platform.

\section{Characterisation and modelling}
\label{sec:characterisation}

\subsection{Index calibration and planar fit}
The wafer fabrication, zinc indiffusion, and ridge dicing procedures of our waveguides follow~\cite{RNCarpenter}. To enable modelling of the waveguide structure, a calibration wafer, processed under the same conditions as the Covesion module (nominal \SI{100}{\nano\metre} Zn, \SI{950}{\celsius}, indiffusion for \SI{1}{\hour}, oxygen ambient), was prism‑coupled at \SI{532}{\nano\metre} (Metricon) to extract three TM planar effective indices and the bulk substrate index. This data was fitted in a planar model, using a commercial fully vectorial optical mode solver (\textsc{FIMMWAVE}, Photon Design), to a Gaussian extraordinary‑index profile,
\begin{equation}
  n_e(x;\lambda_0) = n_{\mathrm{sub}} + \Delta n \exp\!\Big(-\tfrac{x^2}{2\sigma^2}\Big),
  \label{eq:gaussian-profile}
\end{equation}
with $n_{\mathrm{sub}}$ the substrate extraordinary index, $\Delta n$ the peak index contrast, and $\sigma$ the indiffusion depth. Best‑fit parameters are
$\Delta n = 0.0049 \pm 0.0003$ and $\sigma \approx \SI{8.4 \pm 0.2}{\micro\metre}$.

\subsection{Waveguide mode modelling}
\label{sec:waveguide_modelling}
The device is a Zn‑indiffused, MgO‑doped PPLN waveguide (Covesion) with ridge width \SI{10.7}{\micro\metre} (manufacturer specification), height \SI{30}{\micro\metre}, and length \SI{40}{\milli\metre}, pigtailed with PM fibre at \SI{780}{\nano\metre} and \SI{1550}{\nano\metre}. The optical guiding region is defined by zinc indiffusion to a depth of $\sim\!\SI{15}{\micro\metre}$, smaller than the etched ridge height. The on‑module oven provides temperature control. The structure is designed for Type‑0 \SI{1560}{\nano\metre} $\rightarrow$ \SI{780}{\nano\metre} QPM ($e \rightarrow e + e$) and is used here in reverse (SPDC: \SI{780}{\nano\metre} $\rightarrow$ \SI{1560}{\nano\metre}); a nominal poling period of \SI{18.6}{\micro\metre} is required~\cite{RNCarpenter}. For modelling nonlinear interactions a waveguide model was constructed in FIMMWAVE corresponding to the empirical waveguide dimensions.  Material dispersion and its temperature dependence follow the Sellmeier relations of~\cite{Gayer2008}; a small offset aligns $n_e(\SI{532}{\nano\metre}, \SI{25}{\celsius})$ to the measured substrate value ($n_e=2.2208$).

\begin{figure*}[htbp]
  \centering
  \begin{subfigure}[b]{0.33\textwidth}
    \centering
    \includegraphics[width=\linewidth]{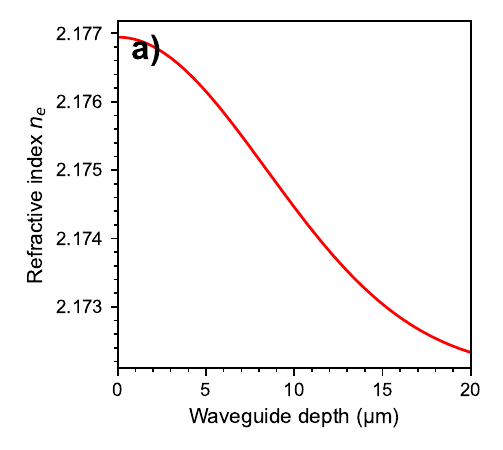}
  \end{subfigure}\hfill
  \begin{subfigure}[b]{0.66\textwidth}
    \centering
    \includegraphics[width=\linewidth]{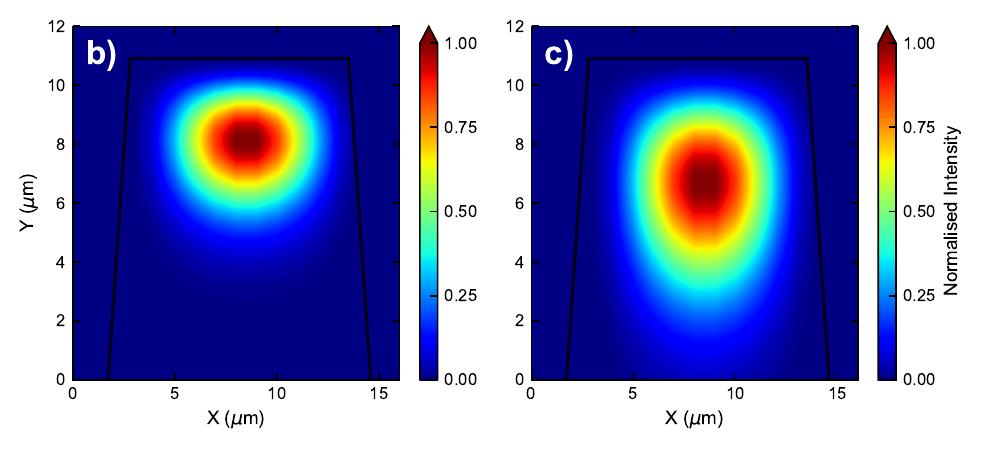}
  \end{subfigure}
  \caption{\textbf{Mode analysis.} (a) Theoretical extraordinary index contrast profile $\Delta n_e(x)$ at \SI{780}{\nano\metre}. (b) Simulated TM$_{00}$ at \SI{780}{\nano\metre}. (c) Simulated TM$_{00}$ at \SI{1560}{\nano\metre}.}
  \label{fig:waveguidemodel}
\end{figure*}

The waveguide is single‑mode at \SI{1560}{\nano\metre} and strongly multimode at \SI{780}{\nano\metre} (pump). Although the module is aligned and qualified using fundamental-mode SHG at \SI{1560}{\nano\metre}, the reverse SPDC process is initiated at \SI{780}{\nano\metre} where the ridge is strongly multimode, so identical mode selectivity cannot be assumed a priori. Simulations indicate both TM$_{00}$ modes are well‑confined with a calculated linear spatial overlap of \SI{77.6}{\percent}, and the \SI{1560}{\nano\metre} mode is near‑circular with excellent fibre compatibility (Fig.~\ref{fig:waveguidemodel}).

\subsection{Second‑harmonic generation (SHG)}
\label{sec:SHG}
\begin{figure}[t]
  \centering
  \begin{subfigure}[t]{0.48\linewidth}
    \centering
    \caption*{(a)}
    \includegraphics[width=\linewidth]{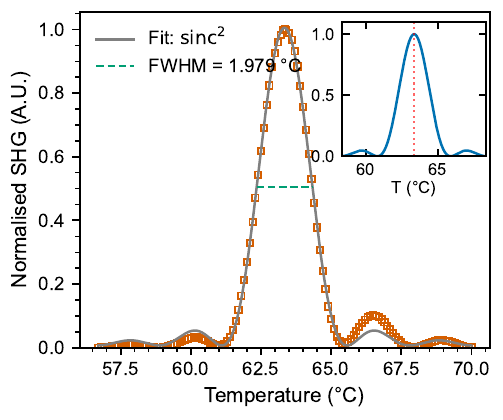}
    \label{fig:shg_2panel:a}
  \end{subfigure}\hfill
  \begin{subfigure}[t]{0.48\linewidth}
    \centering
    \caption*{(b)}
    \includegraphics[width=\linewidth]{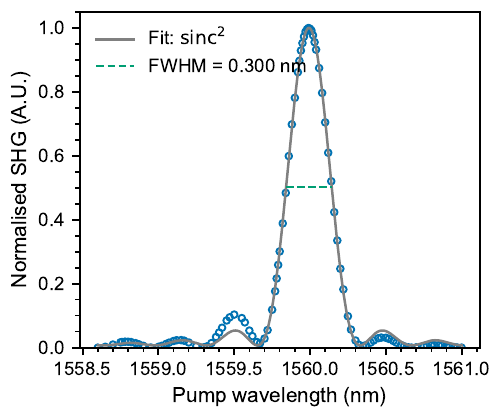}
    \label{fig:shg_2panel:b}
  \end{subfigure}
  \caption{SHG characterisation. (a) Temperature acceptance near $\lambda\approx\SI{1560}{\nano\metre}$. Fits give $\lambda_0=\SI{1559.933}{\nano\metre}$, $\mathrm{FWHM}_{\lambda}=\SI{0.300}{\nano\metre}$ and $T_0=\SI{63.338}{\celsius}$, $\mathrm{FWHM}_T=\SI{1.979}{\celsius}$. (b) Wavelength acceptance at $T\approx\SI{63.34}{\celsius}$: markers, data; grey, $\mathrm{sinc}^2$ fit; green dashed, FWHM segment.}
  \label{fig:shg2_panel}
\end{figure}

We used SHG to locate phase‑matching, exploiting the higher conversion and lower noise than SPDC. Energy conservation holds for both processes ($\omega_p=\omega_s+\omega_i$). The QPM phase‑mismatch is
\begin{equation}
  \Delta k = k_p - k_s - k_i - \frac{2\pi}{\Lambda}, \qquad
  k = \frac{2\pi}{\lambda}\, n_{\mathrm{eff}}(\lambda, T),
  \label{eq:delta-k}
\end{equation}
with $\Lambda$ the nominal poling period (\SI{18.6}  {\micro\metre}). Experimentally, using a narrow‑linewidth C‑band laser (Santec WSL‑110), we measured $T_0=\SI{63.34}{\celsius}$ and $\mathrm{FWHM}_T=\SI{1.979}{\celsius}$ at $\lambda_p=\SI{1560}{\nano\metre}$ (Fig.~\ref{fig:shg2_panel}a, main panel).  A typical  $\mathrm{sinc}^2$ shape is apparent indicating excellent poling homogeneity.

To obtain a theoretical SHG response we evaluated our waveguide model across temperature in \textsc{FIMMWAVE} to obtain $n_{\omega}(T)$ for TM$_{00}$ at \SI{1560}{\nano\metre} and $n_{2\omega}(T)$ for TM$_{00}$ at \SI{780}{\nano\metre}, and evaluated
\begin{equation}
  \Delta k(T) = \frac{4\pi}{\lambda_f}\!\left[n_{2\omega}(T)-n_{\omega}(T)\right] - \frac{2\pi}{\Lambda(T)},
\end{equation}
where $\lambda_f$ is the fundamental vacuum wavelength and $\Lambda(T)=\Lambda_0\,[1+\alpha(T-T_{\mathrm{ref}})]$ includes LN thermal expansion with $\alpha \approx 1.5\times 10^{-5}\,\mathrm{K^{-1}}$~\cite{RNKim}. The theoretical SHG temperature acceptance is shown in Fig.~\ref{fig:shg2_panel}a (inset) with  $\Lambda$  adjusted to \SI{18.57}  {\micro\metre} to align theoretical and experimental peaks. The slight asymmetry in our experimental sidelobes may be due to a small variation in ridge width \cite{RNGrayvariations}. From the simulated group indices at $T_0=\SI{63.34}{\celsius}$,
$n_{g,\omega}(\SI{1560}{\nano\metre})=2.1887$ and
$n_{g,2\omega}(\SI{780}{\nano\metre})=2.2764$, we obtain a SHG spectral bandwidth,
$\Delta\lambda_{\mathrm{FWHM}}\approx\SI{0.307}{\nano\metre}$, matching the measured \SI{0.300}{\nano\metre} (Fig.~\ref{fig:shg2_panel}b). This indicates a highly uniform grating over \SI{40}{\milli\metre} with negligible chirp or area fluctuations.

\subsection{Nonlinear overlap, broadband scaling and idealised brightness}
\label{sec:nonlinear_theory}

For degenerate Type‑0 SPDC, the nonlinear coupling is governed by the spatial overlap of the interacting modes and scales with the overlap integral $E_p E_s^2$, where $E_p(x,y)$ and $E_s(x,y)$ denote the transverse electric field distributions of the pump and degenerate signal/idler modes, respectively. This interaction is conveniently normalised by the effective nonlinear area $A_{\mathrm{eff}}$~\cite{RN305}
\begin{equation}
  A_{\mathrm{eff}}^{(\mathrm{SPDC})} =
  \frac{\left(\iint E_p^2\,\mathrm{d}x\,\mathrm{d}y\right)\!
        \left(\iint E_s^2\,\mathrm{d}x\,\mathrm{d}y\right)^2}
       {\left(\iint E_p E_s^2\,\mathrm{d}x\,\mathrm{d}y\right)^2}.
\end{equation}
where a smaller effective area corresponds to stronger modal confinement and hence increased nonlinear interaction efficiency. We evaluated $A_{\mathrm{eff}}$ numerically using the simulated TM$_{00}$ modes at \SI{780}{\nano\metre} and \SI{1560}{\nano\metre} on the native FIMMWAVE solver grid, thereby capturing exact boundaries and minor field asymmetries. This yields $A_{\mathrm{eff}}=\SI{73.61}{\micro\metre\squared}$.

To obtain the theoretical SPDC bandwidth we used the full dispersion extracted from our FIMMWAVE waveguide model. The wavelength-dependent effective indices $n_{\mathrm{eff}}(\lambda)$ for the TM$_{00}$ mode were used to construct the phase mismatch
\begin{equation}
  \Delta k(\lambda_s) = k_p - k_s - k_i - k_g,
\end{equation}
where $k_j = 2\pi n_{\mathrm{eff}}(\lambda_j)/\lambda_j$, and the idler wavelength $\lambda_i$ is determined by energy conservation. The grating vector $k_g$ was chosen to satisfy $\Delta k = 0$ at degeneracy. The corresponding spectral envelope is given by the standard phase-matching function $\mathrm{sinc}^2(\Delta k L / 2)$.

This approach fully incorporates both material and waveguide dispersion without relying on a low-order Taylor expansion of $\Delta k$. The resulting spectrum was computed numerically, normalised, and its full-width at half-maximum extracted, yielding a theoretical bandwidth of $\sim\!69.6$~nm.

Combining the modal overlap, dispersion-limited bandwidth, and waveguide geometry, the theoretical absolute pair-generation rate for a GVD-limited, degenerate Type-0 SPDC process can be expressed, following \cite{RNRidge}, as
\begin{equation}
  R_{\mathrm{SPDC}} = \frac{d_{\mathrm{eff}}^2 \omega_p^2}{3 \sqrt{2\pi \kappa} \epsilon_0 c^3 n_s^2 n_p A_{\mathrm{eff}}}\,P\,L^{3/2},
  \label{eq:spdc_rate}
\end{equation}
where $\kappa$ is the absolute value of group‑velocity dispersion, $\beta_2$, and $d_{\mathrm{eff}}=2d_{33}/\pi$. For $d_{33}\approx \SI{24}{\pico\metre\per\volt}$~\cite{RNScheeloch}, $A_{\mathrm{eff}}=\SI{73.61}{\micro\metre\squared}$, and $\kappa \approx 9.32\times 10^{-26}\,\mathrm{s^2\,m^{-1}}$ (obtained from the model), the idealised pair-generation rate per unit pump power evaluates to $R_{\mathrm{SPDC}}/P \approx 2.7\times 10^{10}\,\mathrm{pairs\,s^{-1}\,mW^{-1}}$.

%========================
\section{Spontaneous parametric down conversion}
\begin{figure*}[t]
  \centering
  % Row 1 ----------------------------------------------------------
  \begin{subfigure}[t]{0.48\textwidth}
    \centering
    \caption*{(a)}
    \includegraphics[width=\linewidth]{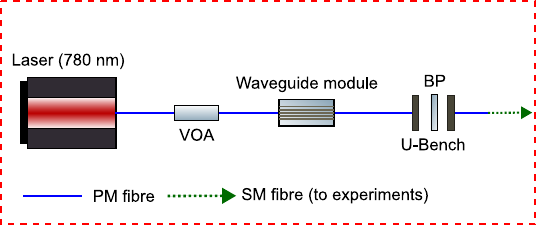}
    \label{fig:spdc_generation}
  \end{subfigure}
  \hfill
  \begin{subfigure}[t]{0.48\textwidth}
    \centering
    \caption*{(b)}
    \includegraphics[width=\linewidth]{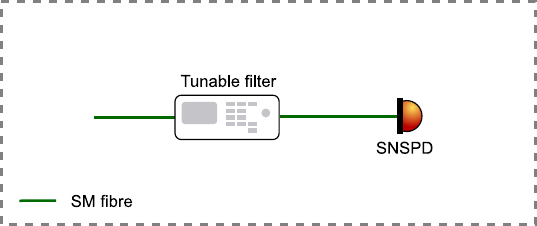}
    \label{fig:spectrum_layout}
  \end{subfigure}

  \vspace{0.1em}

  % Row 2 ----------------------------------------------------------
  \begin{subfigure}[t]{0.48\textwidth}
    \centering
    \caption*{(c)}
    \includegraphics[width=\linewidth]{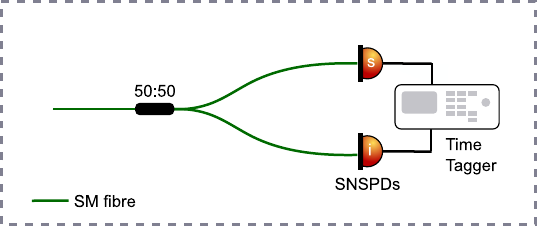}
    \label{fig:car_setup}
  \end{subfigure}
  \hfill
  \begin{subfigure}[t]{0.48\textwidth}
    \centering
    \caption*{(d)}
    \includegraphics[width=\linewidth]{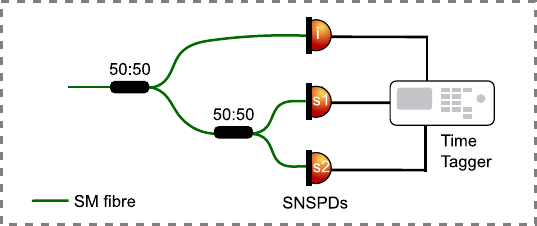}
    \label{fig:g2_setup}
  \end{subfigure}

  \caption{Overview of the SPDC configurations. 
  (a) Type-0 SPDC generation in a fiberised ridge waveguide, the output from which is measured via the following three configurations.  
  (b) Spectral characterisation with a tunable filter. 
  (c) CAR, generated pair rate, and Klyshko efficiency via a 50:50 fibre splitter. 
  (d) Heralded $g^{(2)}(0)$ with two cascaded 50:50 splitters. 
  PM: polarisation-maintaining; SM: single-mode; BP: band-pass; VOA: variable optical attenuator $i$: idler; $s$: signal.}
  \label{fig:main_2x2}
\end{figure*}

%========================
To generate and analyse SPDC, the pigtailed Covesion PPLN ridge module described in Sec.~\ref{sec:waveguide_modelling} was seeded at \SI{780}{\nano\metre}, and the degenerate output at \SI{1560}{\nano\metre} was detected on SNSPDs. The seed was a PM-fibred \SI{780}{\nano\metre} single-mode laser (Thorlabs DBR780PN, nominal \SI{1}{\mega\hertz} linewidth) with a variable optical attenuator (Thorlabs VOA50PM-APC) to set the input power. Residual pump light was rejected using a free-space silicon filter (Thorlabs FBHQ1550)  providing approximately \SI{120}{\decibel} suppression. The SPDC generation stage is shown in Fig.~\ref{fig:main_2x2}(a). To verify pump rejection, we detuned from phase-matching and observed that detector count rates fell to background. The down-converted light was delivered by fibre to three SNSPD channels with specified efficiencies of \SIrange{84}{90}{\percent}, timing jitter \SIrange{27}{31}{\pico\second}, and dead times \SIrange{34}{38}{\nano\second}. Dark counts were negligible under operating conditions. Time tagging used a Swabian TimeTagger with Python-based analysis.

\subsection{SPDC bandwidth measurement}
\begin{figure}[t]
  \centering
  \includegraphics[width=7cm]{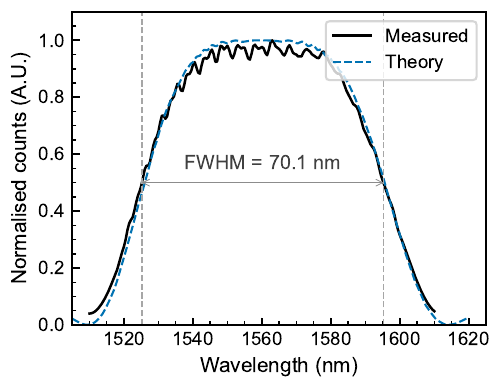}
  \caption{Measured and numerically modelled SPDC marginal spectra, normalised to their respective maxima. The theoretical curve is obtained from the phase-matching function using wavelength-dependent effective indices from the FIMMWAVE model. Close agreement is observed in the full-width at half-maximum.}
  \label{fig:spdcspectrum}
\end{figure}

For bandwidth characterisation, the broadband SPDC output was filtered with a manual tunable filter (Alnair BVF-100) using the configuration of Fig.~\ref{fig:main_2x2}(a and b). The centre wavelength was scanned across the C- and L-bands (\(\approx\)\SIrange{1510}{1610}{\nano\metre}) at the appropriate phase-matching temperature, with a sub-nanometre passband. Normalised counts are shown in Fig.~\ref{fig:spdcspectrum}. The measured marginal spectrum has \(\mathrm{FWHM}\sim\)\SI{70.1}{\nano\metre}, in excellent agreement with the \(\sim\)\SI{69.6}{\nano\metre} expected from the model in Sec.~\ref{sec:nonlinear_theory}. The small residual discrepancy may be attributable to minor deviations from exact degeneracy temperature and the finite spectral resolution of the measurement.

\subsection{CAR and generated photon pair rate}
\label{sec:CARrate}
We quantify pair correlations using the coincidences-to-accidentals ratio (CAR)~\cite{Yadav2022},
\begin{equation}
  \mathrm{CAR} = \frac{C_T}{C_A},
\end{equation}
where \(C_T\) is the zero-delay coincidence count and \(C_A\) is the accidental estimate, obtained as the mean coincidence rate well beyond the coherence time multiplied by the coincidence-window width. Accidentals include background fluorescence, scattering, and residual pump leakage.

\begin{figure}[t]
  \centering
  \begin{subfigure}[t]{0.48\linewidth}
    \centering
    \caption*{(a)}
    \includegraphics[width=\linewidth]{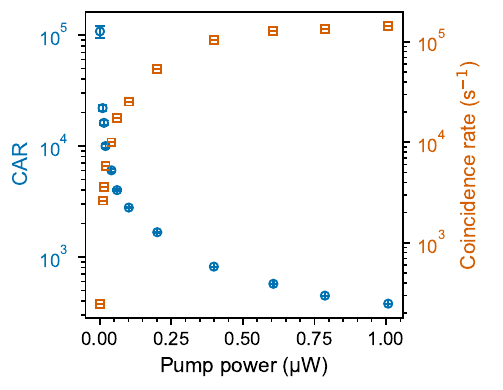}
    \label{fig:carandpairrate:a}
  \end{subfigure}\hfill
  \begin{subfigure}[t]{0.48\linewidth}
    \centering
    \caption*{(b)}
    \includegraphics[width=\linewidth]{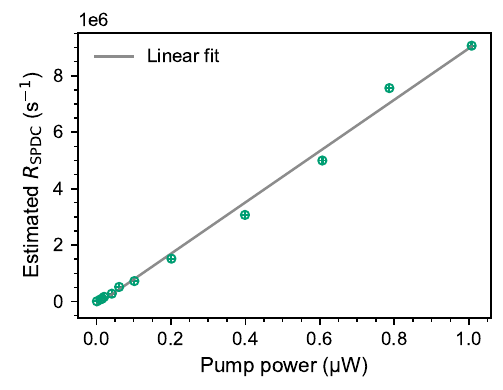}
    \label{fig:carandpairrate:b}
  \end{subfigure}
  \caption{(a) CAR and coincidence rate versus pump power. (b) Estimated SPDC pair rate \(R_{\mathrm{SPDC}}\) versus pump power.}
  \label{fig:carandpairrate}
\end{figure}

Measurements used the setup in Fig.~\ref{fig:main_2x2}(a and c); results are plotted in Fig.~\ref{fig:carandpairrate}(a). Because signal and idler are split probabilistically on a 50:50 coupler, we follow~\cite{RN257} and multiply measured coincidences by two. Coincidence histograms were acquired with \SI{100}{\pico\second} bins. Each SNSPD has \(\sim\)\SI{30}{\pico\second} jitter, so the two-fold peak is well described by a Gaussian-like instrument response function. A fixed integration window was defined as the set of bins for which the coincidence rate exceeded the uniform accidental background by at least a factor of five. This yields a \(\sim\)\SI{600}{\pico\second} symmetric window capturing \(>\)95\% of true pairs while excluding most accidentals; because counts are concentrated in the central three bins, the exact threshold is not critical. At low pump powers we observe \(\mathrm{CAR} > 9\times 10^{4}\) (Fig.~\ref{fig:carandpairrate}a), indicative of strong pair correlations and minimal background photon contributions, consistent with high-quality waveguide SPDC systems~\cite{RN257}.

To estimate the generated pair rate independent of downstream loss, we use a Klyshko-style ratio~\cite{RN285}:
\begin{equation}
  R_{\mathrm{SPDC}} \simeq \frac{R_s^{(\mathrm{adj})}\,R_i^{(\mathrm{adj})}}{R_{s\wedge i}},
\end{equation}
where \(R_s^{(\mathrm{adj})}\) and \(R_i^{(\mathrm{adj})}\) are background-substracted singles rates and \(R_{s\wedge i}\) the coincidence rate (doubled to compensate for probabilistic splitting). Figure~\ref{fig:carandpairrate}(b) shows linear scaling with pump power; the mean normalised brightness is
\(N_{\mathrm{SPDC}}\approx 9.1\times 10^{9}~\mathrm{s^{-1}\,mW^{-1}}\).
This is \(\sim 3\times\) lower than the idealised \(\,2.7\times 10^{10}~\mathrm{pairs\,s^{-1}\,mW^{-1}}\). The waveguide package is built by fibre-launching \SI{1560}{\nano\metre} and optimising fundamental to fundamental SHG to a  \SI{780}{\nano\metre} fibre output. Given the waveguide is multimodal at \SI{780}{\nano\metre}, some of this difference may be attributed to fabrication tolerances affecting the launch of the \SI{780}{\nano\metre} pump in the SPDC process, reducing coupling efficiency.

\subsection{SPDC conversion efficiency and spectral brightness}

We define the SPDC conversion efficiency~\cite{RN257} as
\begin{equation}
  \kappa_{\mathrm{SPDC}} = \frac{R_{\mathrm{SPDC}}}{R_{\mathrm{pump}}},
  \qquad
  R_{\mathrm{pump}} = \frac{P_{\mathrm{pump}}}{h\nu_p} = \frac{P_{\mathrm{pump}}}{h c/\lambda_p},
  \label{eq:kappa_spdc}
\end{equation}
where \(P_{\mathrm{pump}}\) is the transmitted pump power and \(\lambda_p\) the pump wavelength. We obtain
\(\kappa_{\mathrm{SPDC}} \approx 2.3\times 10^{-4}\,\%\).

For spectral brightness we normalise the rate by pump power (mW) and effective detection bandwidth (GHz),
\begin{equation}
  \mathcal{B} = \frac{R_{\mathrm{SPDC}}}{P_{\mathrm{pump}}\, [\mathrm{mW}] \;\Delta\nu\,[\mathrm{GHz}]},
  \qquad
  \Delta\nu \approx \frac{c}{\lambda_0^2}\,\Delta\lambda,
  \label{eq:brightness_spdc}
\end{equation}
with \(\Delta\lambda\) the spectral FWHM and \(\lambda_0\) its centre. 

Using the measured bandwidth \(\Delta\lambda = \SI{70.1}{\nano\metre}\), we obtain 
\(\Delta\nu_{\mathrm{full}} \approx \SI{8660}{\giga\hertz}\), yielding a spectral brightness
\[
\mathcal{B} \approx 1.06\times 10^{6}\,\mathrm{pairs\,s^{-1}\,mW^{-1}\,GHz^{-1}},
\]
equivalently \(1.29\times 10^{8}\,\mathrm{pairs\,s^{-1}\,mW^{-1}\,nm^{-1}}\).

The conversion efficiency compares favourably with prior reports (up to \(4\times 10^{-4}\,\%\)~\cite{Bock2016}); the brightness remains below the \(1.96\times 10^{6}\,\mathrm{pairs\,s^{-1}\,mW^{-1}\,GHz^{-1}}\) reported in~\cite{RNVergyris}, but becomes comparable when normalised for bandwidth.

\subsection{Heralded $g^{(2)}(0)$}
CAR does not directly quantify multi-pair events. We therefore evaluate the heralded second-order correlation using the configuration of Fig.~\ref{fig:main_2x2}(a and d), adopting the unbalanced‑detector form~\cite{RN285} for a nominal 50:50 splitter:
\begin{equation}
  g^{(2)}_{h,50:50}(0) =
  \frac{4\,R_i\,R_{i\wedge s_1\wedge s_2}}{\big(R_{i\wedge s_1}+R_{i\wedge s_2}\big)^2},
\end{equation}
where \(R_{i\wedge s_1\wedge s_2}\) is the triple coincidence between the idler and both signal channels, and \(R_{i\wedge s_1}\), \(R_{i\wedge s_2}\) are the respective idler–signal coincidences. Results are shown in Fig.~\ref{fig:g2}(a). At \(P=\SI{10.225}{\nano\watt}\) we obtain
\(g^{(2)}_{h,50:50}(0) = (5.53 \pm 0.46)\times 10^{-4}\),
well below the single‑photon threshold of \(0.5\)~\cite{RN287}. At the lowest pump (\(\SI{0.535}{\nano\watt}\)) we observe an apparent
\(g^{(2)}_{h,50:50}(0) \approx 3.8\times 10^{-5}\),
but with limited significance (SNR \(\approx 1.2\)) due to \(\sim\!1\) triple event over \(\sim\)8~h. We therefore treat this as an indicative lower bound rather than a robust estimate.
\subsection{Klyshko and heralding efficiency}
\begin{figure}[t]
  \centering
  \begin{subfigure}[t]{0.48\linewidth}
    \centering
    \caption*{(a)}
    \includegraphics[width=\linewidth]{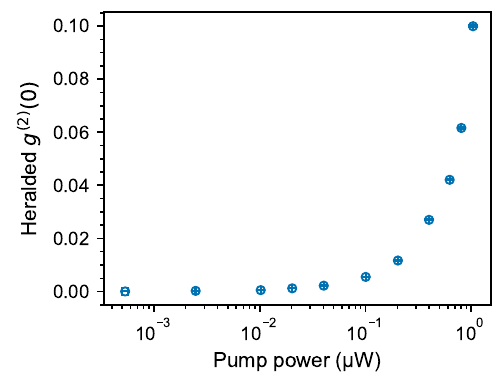}
    \label{fig:g2:a}
  \end{subfigure}\hfill
  \begin{subfigure}[t]{0.48\linewidth}
    \centering
    \caption*{(b)}
    \includegraphics[width=\linewidth]{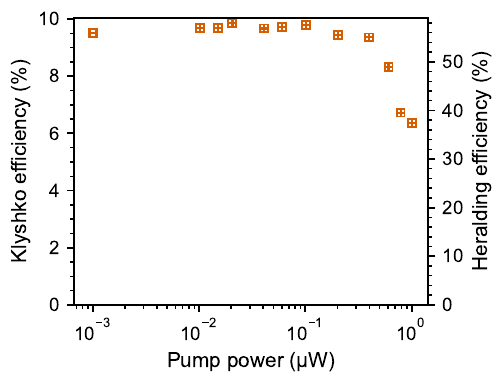}
    \label{fig:g2:b}
  \end{subfigure}
  \caption{(a) Heralded second‑order correlation $g^{(2)}_{h,50:50}(0)$ versus pump
  power. (b) Measured Klyshko and inferred heralding efficiencies after excluding
  probabilistic splitting, detector quantum efficiency and idler‑path optical losses.
  The small error bars in (a) arise from long ($>\SI{1}{\hour}$) integration times.}
  \label{fig:g2}
\end{figure}

To assess the practical heralding performance of the source, including all idler‑path losses (coupling, filtering and detector quantum efficiency), we evaluate the Klyshko efficiency following~\cite{RN285}:
\begin{equation}
    \eta_{k} = \frac{R_{s \wedge i}}{R_{i}},
\end{equation}
where $R_{s \wedge i}$ is the signal--idler coincidence rate and $R_{i}$ is the idler singles rate (corrected for accidentals), using the configuration of Fig.~\ref{fig:main_2x2}(a and c). The choice of heralding arm is arbitrary; slight asymmetries in path loss lead to small differences in the extracted values. Figure~\ref{fig:g2}(b) shows the measured Klyshko efficiencies as a function of pump power, reaching a maximum of \SI{9.870(6)}{\percent} at a pump power for which $g^{(2)}_{h,50:50}(0) \sim 0.001$. Reversing the roles of signal and idler yields a comparable efficiency of \SI{8.910(5)}{\percent}, indicating a balanced and well‑optimised detection setup.

To estimate the internal (module‑level) heralding efficiency of the waveguide, we correct the measured Klyshko efficiency for all idler‑path losses. These include fibre connector losses, the insertion loss of one 50:50 coupler (beyond the ideal \SI{3.01}{\decibel} split) and free‑space filtering losses. Using the measured total idler‑path loss of $7.69 \pm 0.35$~dB, we infer an internal heralding efficiency of \SI{58(5)}{\percent}.

This value naturally includes the loss associated with the \SI{1550}{\nano\metre} output pigtail. To compare directly with free‑space SPDC sources such as~\cite{RN257}, we estimate the corresponding bare‑facet heralding efficiency. Independent classical transmission measurements of a proxy, symmetrically pigtailed device show a total insertion loss of \SI{3.33}{\decibel}. Assuming a propagation loss of approximately \SI{0.1}{\decibel\per\centi\metre} for the \SI{40}{\milli\metre} waveguide, the output pigtail loss is \SI{1.465}{\decibel} (transmission \SI{71.4}{\percent}). Removing this pigtail penalty and applying the \SI{13.2}{\percent} Fresnel loss of a bare $\mathrm{LiNbO_{3}}$--air interface ($n_{s} \approx 2.14$), we infer a bare‑facet internal heralding efficiency of approximately \SI{70.5}{\percent}. This demonstrates that the intrinsic performance of the ridge waveguide is highly competitive with state‑of‑the‑art free‑space platforms (Table~\ref{tab:spdc-sources}), while offering the stability and convenience of permanent fibre integration.

\begin{table*}[t!]
\centering
\caption{Comparison of Type-0 heralded single-photon sources in the telecom band in PPLN waveguides. 
The spectral value for \cite{RN257} was originally reported in per nm. This has been converted to per GHz for consistency. "Platform" indicates metal:indiffusion technology and launch. BW:Bandwidth. Missing values are indicated by ``--''.}
\label{tab:spdc-sources}
\begin{tabular}{lcccccc}
\toprule
\textbf{Reference} & \textbf{Platform} &
\makecell[t]{\textbf{Brightness}\\(pairs/s/mW/GHz)} &
\textbf{CAR} &
\makecell[t]{$\boldsymbol{g_{H}^{(2)}(0)}$} &
\makecell[t]{$\boldsymbol{\eta_{h,\mathrm{internal}}}$\\(\%)} &
\makecell[t]{\textbf{BW}\\(nm)} \\
\midrule
Our source & Zn:fiberised & $1.1\times10^{6}$ & $>90000$ & $<0.001$ & 58 & 70 \\
\cite{Bock2016} & Zn:freespace &  -- & 260000 & 0.001 & 64.1 &  -- \\
\cite{RNVergyris} & --:fiberised &  $2.0\times10^{6}$  & -- & -- & -- &  40 \\
\cite{RNOesterling} & Ti:fiberised & $3.5\times10^{5}$ & -- & -- & 75 & $\sim$45 \\
\cite{Yadav2022} & Zn:fiberised & $2\times10^{5}$ & 668 & -- & -- & 46 \\
\cite{RN257} & Ti:freespace & $2.4\times10^3$ & $\sim$10000 & $<0.003$ & 77.5 & 100 \\
\bottomrule
\end{tabular}
\end{table*}

\section{Conclusion}

We have demonstrated and comprehensively characterised an efficient broadband heralded photon‑pair source based on a fiberised commercial Zn‑indiffused MgO:PPLN ridge waveguide module. Operating in the Type‑0 degenerate SPDC regime ($780~\mathrm{nm} \rightarrow 1560~\mathrm{nm}$), the device produces an emission bandwidth of $\sim\!\SI{70}{\nano\metre}$ spanning the telecom C‑ and L-bands. Such broadband operation makes this architecture well suited to wavelength‑division multiplexed quantum communication schemes.

The source exhibits a strong combination of brightness, purity and heralding performance. We measure a normalised pair‑generation rate of $9.1\times 10^{9}~\mathrm{pairs\,s^{-1}\,mW^{-1}}$ and, after accounting for \SI{7.69 \pm 0.35}{\decibel} of idler‑path loss, infer an internal heralding efficiency of \SI{58(5)}{\percent}. High single‑photon purity is confirmed by a Coincidences‑to‑Accidentals Ratio exceeding $9\times 10^{4}$ and a heralded second‑order correlation of $g^{(2)}(0) < 0.001$. A comparison with other Type‑0 PPLN waveguide sources is provided in Table~\ref{tab:spdc-sources}.

Overall, these results provide a rigorous benchmark for commercial, fibre‑integrated PPLN ridge waveguides as practical single‑photon engines. The combination of high internal heralding efficiency, low multi‑photon noise and broadband emission establishes this platform as a robust and deployable solution for next‑generation, wavelength‑multiplexed quantum networks.
\begin{acknowledgments}
\textbf{Funding.} UK Quantum Technology Hub in Sensing, Imaging and Timing [QuSIT], (EP/Z533166/1); UK Quantum Technology Hub in Quantum Imaging [QuantIC], (EP/T00097X/1); Engineering and Physical Sciences Research Council, (EP/W524621/1).

\textbf{Disclosures.} The authors declare no conflicts of interest.

\textbf{Data availability.} All data supporting this study are available from the University of Southampton repository and will be made publicly available upon publication.
\end{acknowledgments}
\bibliographystyle{apsrev4-2}
\bibliography{references}
\end{document}